\documentclass[prl,aps,twocolumn]{revtex4}

\usepackage{psfrag}
\usepackage{graphicx}
\usepackage{color}
\usepackage{dcolumn}
\usepackage{latexsym,amsfonts}
\usepackage{bm}
\usepackage{amssymb}
\begin{document}

\title{On the time modulation of the K--shell electron capture decay
  rates of H--like heavy ions at GSI experiments}

\author{A. N. Ivanov${^a}$ and P. Kienle$^{b,c}$}
\affiliation{${^a}$Atominstitut der \"Osterreichischen
Universit\"aten, Technische Universit\"at Wien, Wiedner Hauptstrasse
8-10, A-1040 Wien, Austria} \affiliation{${^b}$Stefan Meyer Institut
f\"ur subatomare Physik \"Osterreichische Akademie der Wissenschaften,
Boltzmanngasse 3, A-1090, Wien, Austria} \affiliation{${^c}$Excellence
Cluster Universe Technische Universit\"at M\"unchen, D-85748 Garching,
Germany} \email{ivanov@kph.tuwien.ac.at}

\date{\today}

\begin{abstract}
  According to experimental data at GSI, the rates of the number of
  daughter ions, produced by the nuclear K--shell electron capture
  ($EC$) decays of the H--like heavy ions with one electron in the
  K--shell, such as ${^{140}}{\rm Pr}^{58+}$, ${^{142}}{\rm Pm}^{60+}$
  and ${^{122}}{\rm I}^{52+}$, are modulated in time with periods
  $T_{EC}$ of the order of a few seconds, obeying an $A$--scaling
  $T_{EC} = A/20\,{\rm s}$, where $A$ is the mass number of the mother
  nuclei, and with amplitudes $a^{EC}_d \sim 0.21$. We show that these
  data can be explained in terms of the interference of
  two massive neutrino
  mass--eigenstates. The appearance of the
  interference term is due to overlap of massive neutrino
  mass--eigenstate energies and of the wave functions of the daughter
  ions in two--body decay channels, caused by the energy and momentum
  uncertainties introduced by time differential detection of the
  daughter ions in GSI experiments.  PACS: 12.15.Ff, 13.15.+g,
  23.40.Bw, 26.65.+t
\end{abstract}

\maketitle

\subsection*{Introduction}

Measurements of the K--shell electron capture ($EC$) and positron
($\beta^+$) decay rates of the H--like heavy ions ${^{142}}{\rm
Pm}^{60+}$, ${^{140}}{\rm Pr}^{58+}$, and ${^{122}}{\rm I}^{52+}$ with
one electron in their K--shells have been recently carried out in the
Experimental Storage Ring (ESR) at GSI in Darmstadt
\cite{GSI2}--\cite{GSI5}. The measurements of the rates
$dN^{EC}_d(t)/dt$ of the number $N^{EC}_d(t)$ of daughter ions
${^{142}}{\rm Nd}^{60+}$, ${^{140}}{\rm Ce}^{58+}$ and ${^{122}}{\rm
Te}^{52+}$ showed a time--modulation of exponential decays with
periods $T_{EC} = 7.10(22)\,{\rm s}, 7.06(8)\,{\rm s}$ and
$6.11(3)\,{\rm s}$ and modulation amplitudes $a^{EC}_d = 0.23(4),
0.18(3)$ and $0.22(2)$ for ${^{142}}{\rm Pm}^{60+}$, ${^{140}}{\rm
Pr}^{58+}$ and ${^{122}}{\rm I}^{52+}$ \cite{GSI6},
respectively.

Since the rates of the number of daughter ions are defined by
\begin{eqnarray}\label{label1}
  \frac{dN^{EC}_d(t)}{dt} = \lambda^{EC}_d(t)\, N_m(t),
\end{eqnarray}
where $\lambda^{EC}_d(t)$ is related to the $EC$--decay rate and
$N_m(t)$ is the number of mother ions, the time--modulation of
$dN^{EC}_d(t)/dt$ can be described in terms of a periodic
time--dependence of the $EC$--decay rate $\lambda^{EC}_d(t)$
\cite{GSI2}--\cite{GSI5}
\begin{eqnarray}\label{label2}
  \lambda^{EC}_d(t) = \lambda_{EC}\,(1 + a^{EC}_d \cos(\omega_{EC}t +
\phi_{EC})),
\end{eqnarray}
where $\lambda_{EC}$ is the $EC$--decay constant, $a^{EC}_d$, $T_{EC}
= 2\pi/\omega_{EC}$ and $\phi_{EC}$ are the amplitude, period and
phase of the time--dependent term \cite{GSI2}. Furthermore it was
shown that the $\beta^+$--decay rate of ${^{142}}{\rm Pm}^{60+}$,
measured simultaneously with its modulated $EC$--decay rate, is not
modulated with an amplitude upper limit $a_{\beta^+} < 0.03$
\cite{GSI3}--\cite{GSI5} (see also \cite{GSI6}).

The important property of the periods of the time modulation of the
$EC$--decay rates is their proportionality to the mass number $A$ of
the nucleus of the mother H--like heavy ion. Indeed, the periods
$T_{EC}$ can be described well by the phenomenological formula $T_{EC}
= A/20\,{\rm s}$.  The proportionality of the periods of the
$EC$--decay rates to the mass number $A$ of the mother nuclei and the
absence of the time modulation of the $\beta^+$--decay branch
\cite{GSI3}--\cite{GSI5} can be explained, assuming the interference
of neutrino mass--eigenstates caused by a coherent superposition of
mono--chromatic neutrino mass--eigenstates with electron lepton charge
\cite{Ivanov2,Ivanov4}.

Indeed, nowadays the existence of massive neutrinos, neutrino--flavour
mixing and neutrino oscillations is well established experimentally
and elaborated theoretically \cite{PDG08}.  The observation of the
interference of massive neutrino mass--eigenstates in the $EC$--decay
rates of the H--like heavy ions sheds new light on the important
properties of these states.

\subsection{Amplitudes of  $EC$--decays of H--like heavy ions}

The Hamilton operator ${\rm H}_W(t)$ of the weak interactions,
responsible for the $EC$ and $\beta^+$ decays of the H--like heavy
ions, can be taken in the standard form \cite{Ivanov1}, accounting for
the neutrino--flavour mixing \cite{Ivanov4}. This gives $ {\rm H}_W(t
) = \sum_j U_{ej}{\rm H}^{(j)}_W(t)$, where $U_{ej}$ are matrix
elements of the neutrino mixing matrix $U$ \cite{PDG08} and ${\rm
H}^{(j)}_W(t)$ is defined by
\begin{eqnarray}\label{label3}
\hspace{-0.3in}&&{\rm H}^{(j)}_W(t)
  =\frac{G_F}{\sqrt{2}}V_{ud}\!\!\int\!\!  d^3x
  [\bar{\psi}_n(x)\gamma^{\mu}(1 - g_A\gamma^5) \psi_p(x)]\nonumber\\
\hspace{-0.3in}&& \times [\bar{\psi}_{\nu_j}(x)
  \gamma_{\mu}(1 - \gamma^5)\psi_{e^-}(x)],
\end{eqnarray}
with standard notation \cite{Ivanov1}. The amplitude $A(m \to d +
\nu_e)(t)$ of the $EC$--decay $m \to d + \nu_e$ of a mother H--like
ion $m$ into a daughter ion $d$ and an electron neutrino $\nu_e$ is
defined as a sum of the amplitudes $A(m \to d + \nu_j)(t)$ of the
emission of the massive neutrino mass-eigenstates $m \to d + \nu_j$ as
follows
\begin{eqnarray}\label{label4}
A(m \to d + \nu_e)(t) = \sum_j U_{e j}A(m \to d + \nu_j)(t),
\end{eqnarray}
where the coefficients $U_{ej}$ take into account that the electron
couples to the electron neutrino only $|\nu_e\rangle =
\sum_jU^*_{ej}|\nu_j\rangle$ \cite{PDG08} and $t$ corresponds to the
time of the observation of the decay of the mother ion into the
daughter and electron neutrino state $d + \nu_e$.

According to time--dependent perturbation theory \cite{QM1}, the
partial amplitudes $A(m \to d + \nu_j)(t)$, defined in the rest frame
of the mother ion, are given by
\begin{eqnarray}\label{label5}
\hspace{-0.5in}&&A(m \to d + \nu_j)(t) =\nonumber\\
\hspace{-0.5in} &&= - i\int^t_{-\infty} d\tau\,e^{\,\varepsilon
\tau}\langle
d(\vec{q}_j)\nu_j(\vec{k}_j)|H^{(j)}_W(\tau)|m(\vec{0}\,)\rangle,
\end{eqnarray}
where $\vec{k}_j$ is the 3--momentum of the neutrino mass--eigenstate
$\nu_j$ with mass $m_j$ and $\vec{q}_j$ is the 3--momentum of the
daughter ion produced entangled with the neutrino mass--eigenstate
$\nu_j$. For the regularization of the integral over time we use the
$\varepsilon \to 0$ regularization procedure.  Due to the factor
$e^{\,\varepsilon \tau}$ the weak interaction switches on
adiabatically up to the moment of the production of the mother ion,
followed by its injection into the Experimental Storage Ring (ESR), in
which its decay is observed at the time $t_L = \gamma t$, where $t_L$
is a laboratory time and $\gamma = 1.43$ is a Lorentz factor
\cite{GSI2}, relating times in the laboratory frame and the center of
mass frame. In this connection we recall that at GSI the H-like mother
ions are produced by a fragmentation reaction in a ${^9}{\rm Be}$
target using a $500\,{\rm MeV}$ per nucleon bunched heavy ion beam
with a bunch length of about $300\,{\rm ns}$, separated in a fragment
separator and injected into the ESR after about $500\,{\rm ns}$
transit time at $400\,{\rm MeV}$ per nucleon energy \cite{GSI2}.  The
upper limit $t = t_L/\gamma$ of the integral over time has the meaning
of the time of the observation of the decay of the mother H--like ion
into the final state $d + \nu_e$, which is also the time of the
correlated appearance of the daughter ion $d$, which is actually
observed, and the electron neutrino $\nu_e$ as a coherent
superposition of neutrino mass--eigenstates.

Following \cite{Ivanov1}, we obtain the amplitude of the $EC$--decay,
which is a Gamow--Teller $1^+ \to 0^+$ transition, as a function of
time $t$
\begin{eqnarray}\label{label6}
  \hspace{-0.3in}&&A(m \to d + \nu_e)(t) = -
  \,\delta_{M_F,-\frac{1}{2}}\,\sqrt{3} \sqrt{2 M_m } \nonumber\\
\hspace{-0.3in}&&\times\,{\cal M}_{\rm GT}\,\langle
  \psi^{(Z)}_{1s}\rangle \sum_j U_{ej} \sqrt{2 E_d(\vec{k}_j)
  E_j(\vec{k}_j)}\nonumber\\
   \hspace{-0.3in}&&\times \frac{e^{\,i(\Delta E_j -
  i\varepsilon)t}}{\Delta E_j -
  i\varepsilon}\,\Phi_d(\vec{k}_j + \vec{q}_j),
\end{eqnarray}
where $\Delta E_j = E_d(\vec{q}_j) + E_j(\vec{k}_j) - M_m$ is the
energy difference of the final and initial state, $M_m$ is the mother
ion mass, $E_d(\vec{q}_j)$ and $E_j(\vec{k}_j)$ are the energies of
the daughter ion and massive neutrino $\nu_j$, ${\cal M}_{\rm GT}$ is
the nuclear matrix element of the Gamow--Teller transition $m\to d$
and $\langle \psi^{(Z)}_{1s}\rangle$ is the wave function of the bound
electron in the H--like heavy ion $m$, averaged over the nuclear
density \cite{Ivanov1}.

We take the wave function of the detected daughter ion in the form of
a wave packet. In this case the wave function $\Phi_d(\vec{k}_j +
\vec{q}_j)$ describes a spatial smearing, caused by the differential
time detection of the daughter ions \cite{GSI2}. The time differential
detection of the daughter ions from the $EC$--decays with a time
resolution $\tau_d \simeq 320\,{\rm ms}$ introduces an energy
uncertainty $\delta E_d \sim 2\pi\hbar/\tau_d = 1.29\times
10^{-14}\,{\rm eV}$, thus providing also a smearing of the 3--momenta
of the daughter ions $|\delta \vec{q}_d| \sim 2\pi\hbar/v_d\tau_d =
1.82\times 10^{-14}\,{\rm eV}$, where $v_d$ is the velocity of the
daughter ions, which is in the order of the velocity of the mother
ions $v_d \simeq v_m = 0.71\,c$ in the laboratory frame
\cite{GSI2}. This implies that the wave function $\Phi_d(\vec{k}_j +
\vec{q}_j)$ is peaked in the vicinity of the zero value of the
argument.  In case the wave function of the daughter ion is
approximately a plane wave, the wave function $\Phi_d(\vec{k}_j +
\vec{q}_j)$ is proportional to the $\delta$--function.

Due to energy and momentum conservation in every $EC$--decay channel
$m \to d + \nu_j$ the energy of massive neutrino $\nu_j$ is equal to
\begin{eqnarray}\label{label7}
  E_j(\vec{k}_j) = \frac{M^2_m - M^2_d + m^2_j}{2 M_m},
\end{eqnarray}
which leads to the energy difference of neutrino mass--eigenstates 
\begin{eqnarray}\label{label8}
\omega_{ij} =   E_i(\vec{k}_i)  - E_j(\vec{k}_j) = \frac{\Delta m^2_{ij}}{2 M_m},
\end{eqnarray}
where $\Delta m^2_{ij} = m^2_i - m^2_j$.  Note, that $\omega_{ij}$
determines also the recoil energy difference of the observed daughter
ions.

Since the experimental value of the matrix element $U_{13} =
\sin\theta_{13}\,e^{\,i\,\delta}$, where $\delta$ is a CP--violating
phase \cite{PDG08}, is very close to zero, we set $\theta_{13} = 0$
and below deal with two neutrino mass--eigenstates only with mixing
matrix elements $U_{e 1} = \cos\theta_{12}$ and $U_{e 2} =
\sin\theta_{12}$ \cite{PDG08}.

\subsection*{$EC$--decay rates of H--like heavy ions}

In the following we derive the expression for the $EC$--decay rate of
the H-like heavy ion with a bare daughter ion and a coherent
superposition of electron neutrino mass--eigenstates as the final
state.  The $EC$--decay rate is related to the expression
\begin{eqnarray}\label{label9}
\hspace{-0.3in}&&\lim_{\varepsilon \to
0}\frac{d}{dt}\frac{1}{2}\sum_{M_F}|A(m\to d + \nu_e)(t)|^2 = 3 M_m
|{\cal M}_{\rm GT}|^2 \nonumber\\
\hspace{-0.3in}&&\times\,|\langle \psi^{(Z)}_{1s}\rangle|^2 \Big\{\sum_{j =
1,2}|U_{ej}|^2 2 E_d(\vec{k}_j)E_j(\vec{k}_j)\,2\pi\,\delta(\Delta
E_j)\nonumber\\
\hspace{-0.3in}&&\times |\Phi_d(\vec{k}_j + \vec{q}_j)|^2 + \sum_{i >
j}U^*_{e i}U_{ej}\sqrt{ 2 E_d(\vec{k}_i)E_i(\vec{k}_i)}\nonumber\\
\hspace{-0.3in}&&\times \sqrt{2
E_d(\vec{k}_j)E_j(\vec{k}_j)}\,\Phi^*_d(\vec{k}_i +
\vec{q}_i)\,\Phi_d(\vec{k}_j + \vec{q}_j)\nonumber\\
\hspace{-0.3in}&&\times\,[2\pi\,\delta(\Delta E_i) +
2\pi\,\delta(\Delta E_j)]\,\cos(\omega_{ij}t)\Big\},
\end{eqnarray}
where $\omega_{ij}$ is given by Eq.(\ref{label8}).  The first term in
Eq.(\ref{label10}) is the sum of the two diagonal terms of the
transition probability into the states $d + \nu_1$ and $d + \nu_2$,
describing the incoherent contribution of neutrino mass--eigenstates,
while the second term defines the interference of states $ \nu_i$ and
$ \nu_j$ with $i\neq j$ causing the periodic time
dependence with the frequency $\omega_{ij}$. The interference term,
produced by the coherent contribution of the neutrino
mass--eigenstates, can be only observed due to the energy and momentum
uncertainties, introduced by the time differential detection of the
daughter ions as pointed out above. For the subsequent calculation of
the $EC$--decay rate we can take the massless limit everywhere except
the modulated term $U^*_{e i}U_{ej}\cos(\omega_{ij}t)$. Due to the
momentum uncertainty induced by the time differential detection of the
daughter ions we can set $\vec{k}_i \simeq \vec{k}_j \simeq \vec{k}$
and $\vec{q}_i \simeq \vec{q}_j \simeq \vec{q}$ with the consequence
that $|\Phi_d(\vec{k} + \vec{q}\,)|^2 = V
(2\pi)^3\,\delta^{(3)}(\vec{k} + \vec{q}\,)$, where $V$ is a
normalisation volume and the $\delta$--function describes the
conservation of momentum.

In such an approximation, using the definition of the $EC$--decay rate
\begin{eqnarray}\label{label10}
\hspace{-0.3in}&& \lambda_{EC}(t) = \frac{1}{2M_m V}\int
\frac{d^3q}{(2\pi)^3 2E_d}\frac{d^3k}{(2\pi)^3
2E_{\nu_e}}\nonumber\\
\hspace{-0.3in}&& \times\,\lim_{\varepsilon \to
0}\frac{d}{dt}\frac{1}{2}\sum_{M_F}|A(m\to
d+\nu_e)(t)|^2,
\end{eqnarray}
we obtain the following expression for the time modulated $EC$--decay
rate
\begin{eqnarray}\label{label11}
\lambda_{EC}(t) = \lambda_{EC}(1 +
a_{EC}\,\cos(\omega_{21}t)),
\end{eqnarray}
where $\omega_{21} = \Delta m^2_{21}/2M_m$, $a_{EC} =
\sin2\theta_{12}$ and $\lambda_{EC}$ has been calculated in
\cite{Ivanov1}. In the laboratory frame the $EC$--decay rate is time
modulated with a frequency $\omega_{EC} = \omega_{21}/\gamma$.
Thus, the period $T_{EC}$ of the time modulation is
\begin{eqnarray}\label{label12}
T_{EC} = \frac{2\pi}{\omega_{EC}}= \frac{2\pi \gamma M_m}{\Delta
m^2_{21}},
\end{eqnarray}
where we have taken into account that the period $T_{EC}$ is measured
in the laboratory frame \cite{GSI2}. Since in the massless limit
$\theta_{12} \to 0$, for $m_j \to 0$ we arrive at the
time--independent decay rate $\lambda_{EC}$.

From the experimental data on the periods of the time modulation of
the $EC$--decay rates, the masses $M_m \simeq 931.494\,A$ of mother
ions and Eq.(\ref{label12}) we get the following values for quadratic
mass difference $\Delta m^2_{21}$ of massive neutrinos
\begin{eqnarray}\label{label13}
\hspace{-0.1in}(\Delta m^2_{21})_{\rm GSI} =
\Bigg\{\begin{array}{r@{~,}l} 2.20(7)\times 10^{-4}\,{\rm eV^2} &
{^{142}}{\rm Pm}^{60+} \\ 2.18(3)\times 10^{-4}\,{\rm eV^2} &
{^{140}}{\rm Pr}^{58+}\\ 2.19(1)\times 10^{-4}\,{\rm eV^2} &
{^{122}}{\rm I}^{52+}.
\end{array}
\end{eqnarray}
 The values are equal within their error margins and yield a combined
squared neutrino mass difference $(\Delta m^2_{21})_{\rm GSI} = 2.19
\times 10^{-4}\,{\rm eV^2}$. This confirms the proportionality of the
period of time modulation to the mass number $A$ of the mother nucleus
$T_{EC} = \kappa\,A$, where $\kappa = 4\pi \gamma \hbar M_m/A (\Delta
m^2_{21})_{\rm GSI} = 0.050(4)\,{\rm s}$.

\subsection*{Amplitude of interference term}

The procedure for the calculation of the interference term proposed
above leads to the amplitude $a_{EC} = \sin 2\theta_{12} $ equal to
$a_{EC} \simeq 0.93$, if one takes into account the experimental value
for the mixing angle of two neutrino mass--eigenstates $\theta_{12}
\simeq 34^0$ deduced from solar neutrino experimental data
\cite{PDG08}.

We argue that the experimental amplitude of the time modulation
$a^{EC}_d \sim 0.21$ is not the amplitude of the $EC$--decay rate
$a_{EC}$ of Eq.(\ref{label11}) but presents the amplitude of the time
modulation of the rate of the number of daughter ions. Indeed, in case
the coherence of the contributions of neutrino mass--eigenstates is
retained in all $EC$--decays of the mother ions injected into the ESR,
one should measure the amplitude of the time modulation equal to
$a_{EC}$. However, stochastic processes, affecting a fraction of the
H--like mother ions, can destroy the coherence of the contributions of
neutrino mass--eigenstates and the $EC$--decay rate is defined by the
first term in Eq.(\ref{label9}) only. As a result the $EC$--decay rate
of this fraction of mother ions as well as the rate of the number of
daughter ions do not show a time modulation.  This diminishes
effectively the amplitude of the time modulation of the rate of the
number of daughter ions in Eq.(\ref{label1}), measured by the
experiment \cite{GSI2}--\cite{GSI5}.

For example, the H--like mother ions in the hyperfine ground state
${^A}X^{(Z-1)+}_{F = \frac{1}{2}}$ can be stochastically produced by
the decay of the hyperfine excited states ${^A}X^{(Z-1)+}_{F =
\frac{3}{2}}$ with the emission of a photon with an energy of $\omega
\sim 1\,{\rm eV}$. These preceding electromagnetic decays can destroy
the coherent contribution of neutrino mass---eigenstates to the
$EC$--decay rates of a fraction of mother ions in the process
${^A}X^{(Z-1)+}_{F = \frac{3}{2}} \to m +\gamma \to d + \nu_e +
\gamma$ and lead to the amplitude $a^{EC}_d$ of the time modulation of
the rate of the number of daughter ion smaller than $a_{EC}$. We
propose to test this assertion by studying the $EC$--decays of
He--like heavy ions and the bound-state $\beta^-$--decays of bare ions
with no hyperfine structure.

\subsection*{Conclusion}

We have shown that the experimental data on the time--modulation of
the rates of the number of daughter ions in the $EC$--decays of the
H--like heavy ions, observed at GSI, can be explained by the
interference of massive neutrino mass--eigenstates. 

The necessary condition for the appearance of the interference term in
the $EC$--decay rates is the overlap of the energy levels of neutrino
mass--eigenstates. This is provided by the time differential detection
of the daughter ions with a time resolution $\tau_d \simeq 320\,{\rm
ms}$ leading to an energy uncertainty $\delta E_d \sim 2\pi\hbar/
\tau_d = 1.29\times 10^{-14}\,{\rm eV}$ and a momentum uncertainty
$|\delta \vec{q}_d| \sim 2\pi\hbar/v_d \tau_d = 1.82\times
10^{-14}\,{\rm eV}$ of the daughter ions.

The application of this mechanism to the analysis of time modulation
of the $\beta^+$--decay rates of H--like heavy ions \cite{Ivanov4}
showed the absence of the time modulation due to the broad continuous
neutrino spectrum in accordance with the experimental observation that
the $\beta^+$--decay branch of the H--like ${^{142}}{\rm Pm}^{60+}$
ion decay shows no modulation with an upper limit of the amplitude
$a_{\beta^+} < 0.03$ \cite{GSI3}--\cite{GSI5}.

The value $(\Delta m^2_{21})_{\rm GSI} \simeq 2.19\times 10^{-4}\,{\rm
eV^2}$ is 2.9 times larger than that reported by the KamLAND $(\Delta
m^2_{21})_{\rm KL} = 7.59(21)\times 10^{-5}\,{\rm eV^2}$
\cite{KL08}. A possible solution of this problem in terms of neutrino
mass--corrections, induced by the interaction of massive neutrinos
with strong Coulomb fields of the daughter ions through virtual
$\ell^-W^+$ pair creation, is proposed in \cite{Ivanov3}.

Finally, we emphasize that the ``GSI oscillations'',
i.e. the time modulation or the periodic time--dependence of the rates
of the number of daughter ions of the $EC$--decays of the H--like
mother ions, have no relations to the ``neutrino oscillations'',
e.g. $\nu_e \leftrightarrow \nu_e$, $\nu_e \leftrightarrow \nu_{\mu}$
and $\nu_e \leftrightarrow \nu_{\tau}$. The period of ``neutrino
oscillations'' is expected to be proportional to the neutrino energy
or the $Q$--value of the $EC$--decay. This contradicts the
experimental data showing no dependence of the modulation period on
the $Q$--value but a proportionality to the mass number $A$ of the
mother ion \cite{GSI2}--\cite{GSI5} in agreement with our approach.

Following the publication of the experimental data on the time
modulation of the $EC$--decay rates of the H--like ions ${^{140}}{\rm
Pr}^{58+}$ and ${^{142}}{\rm Pm}^{60+}$ with periods $T_{EC}\simeq
7\,{\rm s}$ \cite{GSI2}, Giunti \cite{Giunti} and Kienert {\it et al.}
\cite{Lindner} proposed that this phenomenon is caused by quantum
beats of two closely spaced mass--eigenstates of the H--like mother
ions with a mass--splitting of order of $10^{-15}\,{\rm eV}$ of
unknown origin.  We notice only that such a mechanism, describing time
modulation of the $EC$--decay rates of the H--like ${^{140}}{\rm
Pr}^{58+}$ and ${^{142}}{\rm Pm}^{60+}$ ions, does not reproduce the
$A$--scaling of the modulation periods, confirmed in the subsequent
experiments on the $EC$--decay rates of the H--like ${^{122}}{\rm
I}^{52+}$ ions \cite{GSI3}--\cite{GSI5}. The
mass--splitting can be attributed to either the nucleus mass or to the
energy level of the bound electron. They would provide either a time
modulation of positron decay rates with a period $T_{\beta^+} =
T_{EC}\simeq 7\,{\rm s}$ or time--independent decay rates for the $EC$
and positron decay, which are both unobserved.

This research was partly supported by the DFG cluster
of excellence "Origin and Structure of rhe Universe" of the Technische
Universit\"at M\"unchen and by the Austrian ``Fonds zur F\"orderung
der Wissenschaftlichenn Forschung'' (FWF) under contract P19487-N16.


\begin{thebibliography}{9}
\bibitem{GSI6} 
This experimental result should be accepted as
``preliminary''.
\end{thebibliography}
\end{document}